**Pareto optimality, economy-effectiveness trade-offs and ion channel degeneracy: Improving population models of neurons**

Peter Jedlicka[a,b,c,*], Alex Bird[a,c,d,#], Hermann Cuntz[c,d,#]

[a] *ICAR3R - Interdisciplinary Centre for 3Rs in Animal Research, Faculty of Medicine, Justus-Liebig-University, Giessen, Germany*
[b] *Institute of Clinical Neuroanatomy, Neuroscience Center, Goethe University, Frankfurt/Main, Germany*
[c] *Frankfurt Institute for Advanced Studies, Frankfurt am Main, Germany*
[d] *Ernst Strüngmann Institute (ESI) for Neuroscience in Cooperation with Max Planck Society, Frankfurt am Main, Germany*
[#]*equally contributed*

**Short title:** Pareto optimality in population neuronal models

**\*Corresponding Author:**

**Peter Jedlicka**

ICAR3R - Interdisciplinary Centre
for 3Rs in Animal Research
Faculty of Medicine
Justus-Liebig-University
Rudolf-Buchheim-Str. 6
D-35392 Giessen
Email: Peter.Jedlicka@informatik.med.uni-giessen.de

**Abstract**

*Nerve cells encounter unavoidable evolutionary trade-offs between multiple tasks. They must consume as little energy as possible (be energy-efficient or economical) but at the same time fulfil their functions (be functionally effective). Neurons displaying best performance for such multi-task trade-offs are said to be Pareto optimal. However, it is not understood how ion channel parameters contribute to the Pareto optimal performance of neurons. Ion channel degeneracy implies that multiple combinations of ion channel parameters can lead to functionally similar neuronal behavior. Therefore, to simulate functional behavior, instead of a single model, neuroscientists often use populations of valid models with distinct ion conductance configurations. This approach is called population (also database or ensemble) modeling. It remains unclear, which ion channel parameters in a vast population of functional models are more likely to be found in the brain. Here we propose that Pareto optimality can serve as a guiding principle for addressing this issue. The Pareto optimum concept can help identify the subpopulations of conductance-based models with ion channel configurations that perform best for the trade-off between economy and functional effectiveness. In this way, the high-dimensional parameter space of neuronal models might be reduced to geometrically simple low-dimensional manifolds. Therefore, Pareto optimality is a promising framework for improving population modeling of neurons and their circuits. We also discuss how Pareto inference might help deduce neuronal functions from high-dimensional Patch-seq data. Furthermore, we hypothesize that Pareto optimality might contribute to our understanding of observed ion channel correlations in neurons.*

**Ion channel degeneracy in population models of neurons**

The landmark studies by Eve Marder and Astrid Prinz have shown that multiple different parameters of ion channels can generate similar activity both at the level of single cells (Prinz et al., 2003) as well as neural circuits (Prinz et al., 2004; Marder and Goaillard, 2006). This multiple-to-one mapping between combinations of ion channel parameters and cell or circuit phenotypes has been termed ion channel degeneracy (Drion et al., 2015) or non-uniqueness (Druckmann et al., 2007; Prinz, 2017). Degeneracy (Edelman and Gally, 2001) is present at all scales of the brain (Fig. 1). Its importance for the flexibility and robustness of brain functions has been increasingly acknowledged in recent years (for reviews see Rathour and Narayanan, 2019; Goaillard and Marder, 2021; Kamaleddin, 2022). Accordingly, ion channel degeneracy has been linked to the flexibility (Drion et al., 2015) and robustness of neuronal behavior (O'Leary, 2018; Onasch and Gjorgjieva, 2020; Goaillard and Marder, 2021).

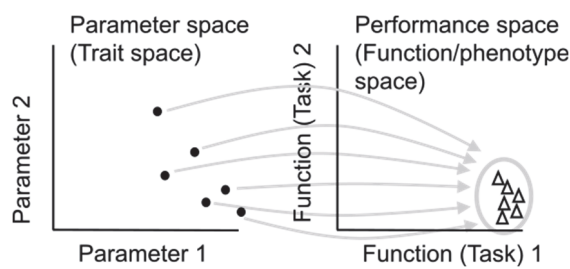

*Figure 1 Degeneracy (partial redundancy) in the parameter space of biological systems (e.g., neurons with ion channels). Multiple disparate parameter configurations in the parameter (trait) space (e.g., ion conductance space) can lead to similar functional phenotypes optimized for a given task 1 (e.g., dendritic computation). In degenerate systems such as our brain, there is a multiple-to-one mapping between the parameter space and the phenotype space at all scales including the scale of ion channels and nerve cells (and their circuits). Each point (triangle) may represent a single neuron in a multidimensional parameter (performance) space. The schematic shows a 2D space but in real systems, parameter and performance space can have different numbers of dimensions (see also Fig. 8). The degeneracy and Pareto optimality concepts can be applied to any number of dimensions. For a 3D-version of a similar schematic see e.g., figure 4 in Mishra and Narayanan (2021). Throughout this article, we consider "parameter space" and "trait space" synonyms. Similarly, we consider "performance space", "functional space", "phenotype space" and "output space" synonyms. This applies also to (neuronal) "tasks", "objectives" and "functions".*

Several groups have adopted computational insights and methods of Marder and colleagues to explore ion channel degeneracy in different types of neurons (e.g., (Keren et al., 2005; Achard and De Schutter, 2006; Druckmann et al., 2007; Günay et al., 2008; Doloc-Mihu and Calabrese, 2011; Sekulic et al., 2014; Neymotin et al., 2017; Migliore et al., 2018; Mittal and Narayanan, 2018; Nandi et al., 2020; Allam et al., 2021). This approach has been successfully used even outside of neuroscience, for example in heart cell physiology (Britton et al., 2013). It has been called population- (Marder and Taylor, 2011; Britton et al., 2013; Allam et al., 2021) or database- (Günay et al., 2008; Doloc-Mihu and Calabrese, 2011) or ensemble-modeling (Sekulic et al., 2014). Population-based computer models have provided a better understanding of cell-to-cell as well as animal-to-animal variability of electrophysiological and ion channel expression data (Marder and Goaillard, 2006; Marder, 2011; Marder and Taylor, 2011). Instead of a "one-size-fits-all" approach in which a computer model simulates average properties of a nerve cell or heart muscle cell, the population-based approach constructs and validates large populations of realistic cellular models that differ in their ion channel configurations and reflect the variability of experimental data (Taylor et al., 2009; Britton et al., 2013). Recent work has also shown that such population models may allow pharmacological predictions *in silico*, thus complementing, and partially replacing animal experiments (Passini et al., 2017; Allam et al., 2021).

Ion channel degeneracy applies not only to intrinsic cellular properties but also to extrinsic synaptic properties (Prinz et al., 2004). Already the first landmark studies have shown that many disparate configurations of synaptic and intrinsic conductances are able to generate similar neuronal behavior as well as similar (functional) network behavior (Prinz et al., 2004; see also Goaillard et al., 2009; Zhao and Golowasch, 2012; Knox et al., 2018; Deistler et al., 2022; Medlock et al., 2022). Therefore, although in this article we focus on degeneracy of intrinsic ion channels at the cellular level, the concepts of degeneracy and Pareto optimality (Pallasdies et al., 2021) can be extended also to extrinsic synaptic channels and to the level of neuronal circuits.

**The problem of the large and complex parameter space of functional models**

Given the importance and the success of the population modeling approach, it would be desirable to further improve its predictive power. This would facilitate clinically relevant predictions about the role of ion channels in neurological diseases with known ion channel expression correlates, such as epilepsy. However, to achieve this, we need to find a solution to one particularly problematic issue of population modeling. Due to ion channel degeneracy, one can obtain similar neuronal computational, functional, and electrophysiological properties with widely different parameter combinations in any given neuronal biophysical model. The problematic issue is that it is unclear which compositions of ion channels and their parameters, all of which generate realistic (functional) electrophysiological behavior, are in reality preferred by evolutionary selection. In other words, it is not understood, which ion channel configurations in a vast population of valid models are more likely to be found in the brain. We know that often there is a degenerate multiple-to-one mapping between parameter (or trait) space of ion channels and phenotype (or performance/function) space of neurons (Fig. 1). However, we lack a theoretical framework to fully constrain this mapping in a biologically realistic manner.

Thus, it is an unresolved question whether naturally occurring configurations of neuronal parameters occupy a large or a restricted subspace in the large, theoretically possible parameter space. The complex shape of the valid parameter space has been explored before (e.g., Achard and De Schutter, 2006; Taylor et al., 2006; Schneider et al., 2021). However, there is a need for universal guiding principles that would further constrain the shape of the parameter space to those models that most likely represent real neurons found in nature. Such a principle would help address the following questions. Are naturally occurring instances of real neurons (and their circuits) confined to low-dimensional manifolds or rather scattered widely over the entire parameter space (Szekely et al., 2013)? In case a neuron type has n ion channel parameters, each instance of the neuron can be represented as a point in an n-dimensional parameter space. Can complex n-dimensional conductance spaces (Taylor et al., 2006) be reduced to low-dimensional subspaces? If real (naturally occurring) configurations of parameters were restricted to low-dimensional manifolds (Mishra and Narayanan, 2021), it would greatly enhance our understanding of neuronal systems. It would potentially allow us to infer most unknown parameters from a small subset of known parameters (Szekely et al., 2013).

Intriguingly, ten years ago, in their pioneering research, Uri Alon and colleagues started using Pareto optimality to show that evolution selects phenotypes that are located in low-dimensional manifolds (e.g., lines, triangles) of parameter space (Shoval et al., 2012). Pareto theory predicts such a low-dimensional geometry of parameter space would be found in nature. The framework of Pareto optimality explains it as a consequence of evolutionary optimization of the phenotypes for their multiple tasks (functions) and making optimal trade-offs between the tasks (Shoval et al., 2012; Szekely et al., 2013).

**Multi-objective Pareto optimality as a geometrically elegant solution for simplifying parameter space**

Evolutionary restriction of a complex parameter space to a simpler subspace or a low-dimensional manifold is per se a plausible and realistic assumption (Fig. 2). However, it remains unclear what additional principle can help us in practice to reduce the parameter space of degenerate ion channels in populations of neuron models. Ideally, such a principle would allow us to identify or at least approximate the shape of the subspaces or manifolds selected by evolution. Pareto optimality linked to evolutionary trade-offs (Alon, 2020) is a promising candidate for such a general and at the same time practical principle.

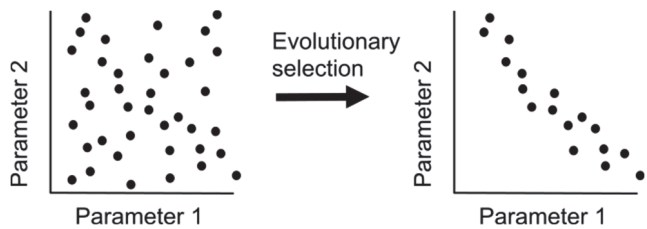

*Figure 2 Evolutionary selection based on trade-offs between multiple tasks can remove suboptimal points from (ion channel) parameter space. Each (neuronal) phenotype can be seen as a point in a 2- or n-dimensional parameter space. There are two possibilities for the geometry of parameter space: Left: The (ion conductance) parameters that contribute to the function(s) of a neuron fill the entire parameter space. Right: Parameters occurring in nature (in real neurons) are restricted to a small subspace (or a curve) of the parameter space because evolution removes inefficient or ineffective parameter configurations. This can be generalized to any number of dimensions. Evolution can confine high-dimensional parameter space to a low dimensional manifold (Szekely and Alon, 2013).*

Usually a neuron (in fact any artificial or a biological system) has to fulfil more than one task at the same time. For example, it must generate functional electrical behavior (e.g. dendritic spikes and/or somatic bursting) and/or maintain its stable (fast or slow) firing and at the same time expend as little energy as possible. In addition, a neuron often has to be robust against perturbations and/or flexible enough to respond to a wide input range. Nevertheless, typically its performance cannot be optimal for all of these separate tasks. Thus, neurons (and their circuits) face a fundamental optimization problem of finding an optimal trade-off among multiple objectives (Laughlin and Sejnowski, 2003; Del Giudice and Crespi, 2018). Evolutionary multi-objective optimization can be seen as a tug-of-war game between many tasks (Hadsell et al., 2020) pulling neurons towards Pareto optimal multi-task solutions.

A key evolutionary hypothesis is that if multiple competing tasks affect the fitness of a phenotype then evolution will select individual phenotypes with optimal performance for, potentially different, combinations of those tasks (for trade-offs between them, see Alon, 2020). In such a case, Pareto optimality may help identify the sets of neurons, for which evolution solved the multi-task optimization problem. By definition, the performance of such a set of Pareto optimal neurons cannot be improved for any task without decreasing their performance for some other task. This means that no other plausible neuron can dominate a Pareto optimal neuron by outperforming it at all tasks simultaneously. The set of Pareto optimal neurons form a so-called Pareto front in task or function space. In contrast to the parameter space, the space of plausible neuronal functionality is typically restricted, with the Pareto front forming part of the boundary manifold between plausible and implausible regions. A neuron belongs to a Pareto front if and only if the following condition is satisfied (Avena-Koenigsberger et al., 2014): For any other distinct neuron in the population, there must exist at least one task at which the Pareto front neuron is strictly better. We can examine such Pareto front sets of neurons first in performance (Fig. 3, 4) and then in parameter space (Fig. 5).

Let us assume that neurons need to optimize their biophysical design (ion channel parameters) for two tasks simultaneously, for example low energy expenditure and dendritic computation. If one neuron (e.g., N4 in Fig. 3) underperforms another neuron (e.g. N3) at both tasks simultaneously then it does not belong to the Pareto front and has been likely eliminated by evolutionary selection. If we remove all neurons that are outperformed (i.e. dominated) by other neurons, we get the Pareto front. Thus, the set of neurons that cannot be outperformed concurrently at both tasks (objectives) is the Pareto-front-set. The neurons (phenotypes) that achieve peak performance for one objective (N1 and N2) are called archetypes (Shoval et al., 2012; Alon, 2020). Archetypes are phenotypes with best combinations of parameters (traits) for given tasks. N1 is the neuronal phenotype with such a combination of ion channel parameters that leads to the best performance in dendritic computation. In contrast, N2's ion channel parameters support its best performance in energy-efficiency (i.e. economy: defined e.g., as low ATP consumption per spike, see below). N5 represents an impossible configuration of parameters with effective dendritic computation, but which is not achievable at such a high energy-efficiency.

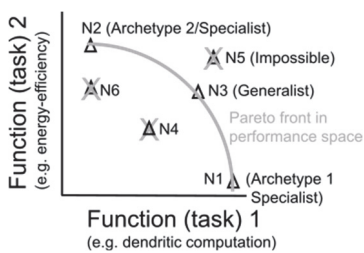

*Figure 3 Neuronal phenotypes that cannot be outperformed in all tasks simultaneously are Pareto optimal and lie on the Pareto front. Neuronal phenotypes, which correspond to different ion channel configurations, can be plotted in performance space based on their performance in 2 (or n) tasks. N1, N2 and N3 neurons outperform N4 and N6 in at least one task. Best phenotypes for a given task (1 or 2) are referred to as archetypes (1 or 2, respectively). Pareto optimal neurons that are close to archetypes are called task specialists. Neurons that are in the middle can be called generalists. Based on Alon (2020).*

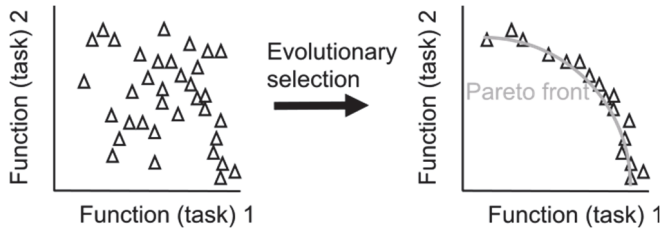

*Figure 4 Evolutionary trade-offs between different tasks reduce performance space to a Pareto front. The key hypothesis is that evolutionary selection based on multi-task trade-offs removed suboptimal neuronal phenotypes from performance space and greatly simplified performance (and the corresponding parameter) space. Based on Alon (2020). Caveat: If measured neuronal phenotypes do not lie on a Pareto front predicted by a Pareto analysis, this could mean that the analysis neglected some important tasks or that neurons are not Pareto optimal for the studied tasks (see Pallasdies et al., 2021)*

Strikingly, Uri Alon and colleagues have shown that the Pareto optimality principle can elegantly simplify the geometry of parameter space, in fact more clearly than it can the geometry of performance space (Shoval et al., 2012; Sheftel et al., 2013). Two tasks or objectives push neurons with an optimal trade-off to a low-dimensional subregion in parameter space that corresponds to a Pareto front. Interestingly, irrespective of the number of parameters (traits), in the case of two tasks, the Pareto front must have a shape of a line connecting two archetypes (Fig. 5A). The reason is that any neuron that is not located on the line (e.g., N4 in Fig. 5B) is necessarily further away from the two archetypes (N1, N2) than any neuron on the line (e.g., N3). This is the case in parameter space, but not in performance space. Performance landscape can be visualized in parameter space by drawing performance contours (Fig. 5A) around archetypes (Shoval et al., 2012). Archetypes represent peaks of performance for single tasks. Importantly, performance decreases with a growing distance from archetypes (in parameter space). For geometrical reasons, each neuron belonging to the Pareto front line must have a lower total distance to both archetypes than any neuron outside the Pareto front line. The lower distance to both archetypes means higher simultaneous performance in both tasks. Pareto optimal neurons that are near the ends of the Pareto line segment are specialists for one of the two tasks, while those in between are generalists for both tasks.

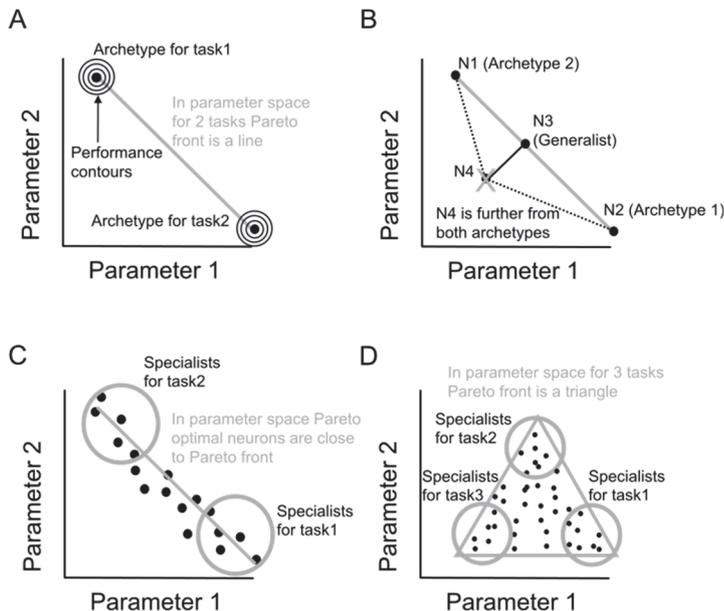

*Figure 5 Trade-offs between 2 or 3 tasks reduce parameter space to low-dimensional Pareto fronts in the form of a line or a triangle, respectively. A) Archetypes are located at peaks of a task performance. Contours indicate a monotonic drop in performance for locations further away from archetypes. B) Nonoptimal neurons (N4) are more distant from archetypes than neurons on the Pareto front (N1,2,3), which is the line between archetypes (N1,2). C), D) Hypothetical neurons that are concurrently optimized for 2 or 3 objectives (e.g., dendritic computation, its energy-efficiency and robustness) would be found on a line segment or inside a triangle with specialists near the 2 or 3 vertices (archetypes), respectively. Based on Alon (2020).*

Similar geometrical reasoning can show that, for three tasks, the Pareto front must have the shape of a triangle in parameter space (Fig. 5D). Intriguingly, this is true irrespective of the number of dimensions of the parameter space. Again, whereas specialists, optimal for a single task, concentrate close to one of the three corners (archetypes), generalists, optimal for combinations of three tasks, occupy the region in the middle. This can be extrapolated to any number of tasks. For four tasks, Pareto front is a tetrahedron with four corners or vertices (archetypes). For n tasks, Pareto front is a polytope with n vertices or corners (Sheftel et al., 2013; Alon, 2020).

Importantly, these geometrical insights hold under three assumptions about task performance (Shoval et al., 2012; Sheftel et al., 2013): (1) Performance decays monotonically with increasing remoteness from the peak (archetype), (2) there is one point representing a global peak, (3) all performances decay with the same metric distance from corresponding peaks. However, even after violating these conditions, approximate Pareto fronts with relatively simple shapes can still emerge, still having archetypes as vertices. The vertices are then connected by mildly curved instead of straight lines, still corresponding to where the performance contours for different tasks lie tangentially to one another (Shoval et al., 2012; Sheftel et al., 2013).

**Evolutionary trade-offs between functionality (effectiveness) and energy-efficiency (economy) of neurons**

How could we take advantage of the above geometrical principles (Shoval et al., 2012; Alon, 2020) derived from Pareto optimality? How can we apply them to reduce the parameter space of degenerate neuronal models? Ion channel degeneracy implies multiple valid solutions (i.e. a population of solutions) for realistic voltage traces. The valid solutions can fill large and distributed regions of parameter space (see e.g., Achard and De Schutter, 2006; Taylor et al., 2006; Schneider et al., 2021). However, although being valid with respect to reproducing voltage traces, individual models in a population differ regarding their optimality for additional functions or objectives. One such important additional objective is energy-efficiency (i.e. economy or low ATP expenditure).

It is well established that to achieve their computational goals, brain circuits and nerve cells consume large amounts of energy (Attwell and Laughlin, 2001; Niven, 2016; Levy and Calvert, 2021). Therefore, their anatomical and physiological properties are likely to be optimized for a fundamental trade-off between function-effectiveness (i.e. effectiveness) and energy-efficiency (i.e. economy). This would be in close agreement with the Pareto optimality concept although in most neuroscience studies it has not been named as such (for a recent review, see Pallasdies et al., 2021). It has been increasingly recognized that evolution has optimized neurons and their circuits for best simultaneous performance in terms of functional effectiveness and economy (Laughlin and Sejnowski, 2003; Sterling and Laughlin, 2015; Pallasdies et al., 2021). Thus, Pareto optimality for the trade-off between effectiveness and economy, which has been used to better understand nonneuronal systems (Szekely et al., 2013) can also be applied explicitly to neurons as a general principle. In line with this, also in neural information theory, accumulating computational and experimental evidence shows that neurons are not optimized for processing maximum amounts of information but rather maximum amounts of information per energy cost (Balasubramanian et al., 2001; Levy and Baxter, 2002; Goldman, 2004; Sengupta et al., 2013c; Yu and Yu, 2017; Stone, 2018). Although not explicitly using Pareto theory, several studies have indicated that both extrinsic (synaptic) as well as intrinsic ion channel properties of neurons are concurrently optimal for high energy-efficiency (economy) and effective information processing and its biophysical implementation (e.g., Alle et al., 2009; Carter and Bean, 2009; Hasenstaub et al., 2010; Sengupta et al., 2010, 2013a, 2013b; Harris et al., 2015; Yu et al., 2018; Mahajan and Nadkarni, 2020; Schug et al., 2021).

Intriguingly, Remme and colleagues have recently used conductance-based modeling to show explicitly that neurons in the medial superior olive (MSO) are close to the Pareto-front set of models in performance space (Remme et al., 2018). The Pareto front neurons were optimal for a two-task trade-off between energy expenditure and a well-known MSO neuronal function, namely detection of temporal coincidence in input signals (Fig. 6). The temporal coincidence detection is crucial for the computation of the direction of sound source in the MSO. Remarkably, a default model (open star in Fig. 6) that was experimentally well constrained for biophysical (ion channel) and morphological parameters exhibited high energy-efficiency and at the same time high functional (computational) effectiveness. This pioneering study is probably the first publication that has explicitly applied the Pareto optimality theory to an effectiveness-economy trade-off in a conductance-based neuronal model (Fig. 6). The modeling work has indicated that neurons minimize their energy costs related to their ion channel parameters as long as their computational function remains intact (Remme et al., 2018).

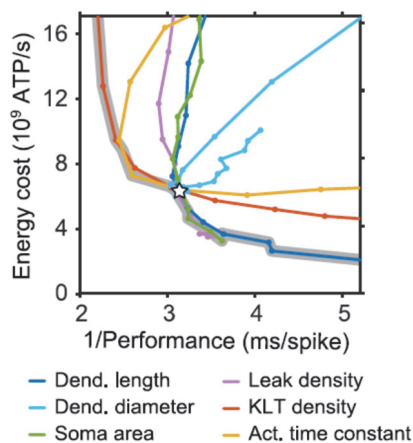

*Figure 6 Simulated neurons in the medial superior olive (MSO) are Pareto optimal for a 2-task trade-off between functional effectiveness and energy-efficiency (economy). The figure shows a performance space for 2 tasks simulated in MSO neuron models. Simulations were run in conductance-based neuronal models with simplified morphologies. Performance in dendritic computation (task 1, x-axis) in the form of temporal coincidence detection for inputs (underlying sound localization in the MSO) is plotted against performance with respect to energy-efficiency (task 2, y-axis). The grey line indicates the Pareto front with models that are optimal for the trade-off between the 2 tasks. Colored curves show models with varied morphological and biophysical (ion channel) parameters. Energy cost was estimated using a standard approach for converting ionic currents into ATP. KLT: low threshold potassium current. Note that a model, which was well constrained by experimental data (open star), is close to the Pareto front and displays strong performance in coincidence detection and high energy-efficiency (i.e. low energy cost). Reproduced from Remme et al. (2018), licensed under Creative Commons Attribution License. https://doi.org/10.1371/journal.pcbi.1006612.g004*

Interestingly, the morphological parameter space of neuronal dendrites and axons can be understood to some extent separately from the biophysical parameter space (Wen and Chklovskii, 2008; Papo et al., 2014). A morphological modeling study by Cuntz et al. (2007) showed that dendritic trees of neurons search for a compromise between two costs: cable length and conduction time. Minimization of cable length can be seen as minimization of energy cost (i.e. maximization of efficient wiring), whereas minimization of conduction time can be seen as effective signal propagation (Chklovskii, 2004; Cuntz et al., 2010; Avena-Koenigsberger et al., 2014, see also Fig. 1 in Pallasdies et al., 2021). A more recent publication has confirmed and extended findings of Cuntz et al. (2007, 2010) showing that by implementing optimal solutions for the trade-off between cable length and conduction time, dendritic trees achieve Pareto optimality (Chandrasekhar and Navlakha, 2019). Thus, although many open questions about the relationship between morphology and biophysics in neuronal models remain (see e.g., Stiefel and Torben-Nielsen, 2014), these studies indicate that dendrite morphology is well constrained by optimal wiring alone. Similar observations have been made for axonal connections (Chen et al., 2006; Budd et al., 2010; Avena-Koenigsberger et al., 2014; Ollé-Vila et al., 2020; Ma et al., 2021; but see also Stiefel et al., 2013). Therefore, in this article we focus mostly on the conductance space and do not discuss the morphological space of population neuronal models and its impact on the variability and robustness of electrophysiological behavior (for this topic see e.g., Beining et al., 2017; Cuntz et al., 2021) or their potential interactions.

**Effectiveness-economy trade-offs may simplify conductance space of population models of neurons**

So far the Pareto optimality theory with a focus on effectiveness-economy trade-offs has not yet been explicitly used to address directly the problem of ion channel degeneracy in population models of neurons (but see also Druckmann et al., 2007, and the discussion below). We argue that multi-objective Pareto optimality, which can be estimated for trade-offs between known (but to some extent also unknown, see Pareto task inference below) computational functions and energy-efficiency could be a very fruitful theoretical framework for improving population models of neurons. Energy costs of individual conductance-based models in populations of valid single- or multi-compartmental models are usually not considered. However, an estimation of energy costs for a given ion channel configuration is relatively straightforward and easy to implement (see e.g., Alle et al., 2009; Remme et al., 2018). Currents flowing through ion conductances in compartmental models can be collected and converted to ATP costs. The estimated ATP amount is proportional to energy consumed by ATP-driven pumps, which maintain transmembrane concentration gradients of sodium, potassium and calcium ions. ATP calculated in this way is a good approximation for the energy costs of conductance-based models. Therefore, in the context of the Pareto optimality framework, we suggest that, whenever possible, energy-efficiency should be added as an additional objective to inform the search for most realistic configurations of ion channel parameters. Moreover, we propose to extend the objectives by evaluating the models not only by their voltage trace features but also by their performance in well-defined computational functions. Examples would include for instance coincidence detection of MSO neurons (Remme et al., 2018) or coincidence detection in the form of BAC firing of cortical layer 5 pyramidal neurons (Larkum et al., 1999; Hay et al., 2011), orientation tuning of layer 2 pyramidal neurons (Goetz et al., 2021) or spatial tuning of grid cells (Schmidt-Hieber and Häusser, 2013) and place cells (Seenivasan and Narayanan, 2020), or pattern separation in dentate granule cells (Madar et al., 2019), or motion detection in direction-selective T4 neurons (Groschner et al., 2022).

It is important to note that feature-based multi-objective optimization has already been established before as an extremely helpful method for tuning compartmental models (Druckmann et al., 2007). In a landmark study, Druckmann, Segev and colleagues used genetic algorithms to optimize multiple objectives in the form of selected features of experimental voltage traces such as frequency, timing or width of action potentials (Druckmann et al., 2007). This approach has been very successful as a feature-based optimization tool (Druckmann et al., 2007; Van Geit et al., 2008, 2016; Hay et al., 2011;

Neymotin et al., 2017; Mäki-Marttunen et al., 2018; Migliore et al., 2018; Nandi et al., 2020). We propose extrapolating multi-objective optimization from spiking features to additional neuronal functions and their energy costs. This would extend population neuronal modeling beyond reproducing electrophysiological features toward capturing evolutionary trade-offs between physiological functions and their energy costs. In other words, conceptually we suggest using multi-objective trade-offs in a more general evolutionary context (cf. Hausser et al., 2019). Furthermore, we propose using the above-described geometrical principles (Shoval et al., 2012; Alon, 2020) for exploring whether it is possible to reduce high-dimensional parameter space of ion conductances to low-dimensional manifolds. Importantly, multi-task Pareto optimization or selection of neuronal models based on function-economy trade-offs can be combined with or complement standard multi-objective optimization based on trade-offs between voltage features. The hope is that such a complementary use of the two non-exclusive overlapping approaches (Druckmann et al., 2007; Remme et al., 2018) can further constrain the parameter space in a biologically realistic manner.

Indeed, recent modeling efforts suggest that such an extended multi-objective optimality framework could be a promising approach for tackling ion channel degeneracy in populations of conductance-based models. New computational studies employing population models of neurons or neuronal circuits have provided further hints that effectiveness-economy trade-offs contribute to a better understanding of the complex parameter space and its degeneracy. One study (Bast and Oberlaender, 2021) used, in a first step, the classical feature-based multi-objective optimization (Druckmann et al., 2007; Hay et al., 2011) to generate multi-compartmental population models for layer 5 pyramidal tract neurons (L5 PCs). Then, in a second step, the feature-optimized models were further analyzed for their energy-efficiency using the above-mentioned ATP estimation approach based on monitoring ionic currents (Remme et al., 2018). Not surprisingly, the population models of L5 PCs showed extensive degeneracy of ion channels since nonlinear dendritic computation emerged from a large range of ion channel configurations. Notably, their computational analysis identified models that were efficient in terms of energy as well as effective in terms of dendritic computation. The nonlinear dendritic computation was assessed based on BAC firing (Larkum et al., 1999; Larkum, 2013) and dendritic calcium spikes, which are thought to be important for conscious perception (Aru et al., 2020; Suzuki and Larkum, 2020; Takahashi et al., 2020).

Curiously, the L5 PC models with energy-efficient dendritic computation displayed a low expression of fast non-inactivating potassium channels and high-voltage activated calcium channels in the dendritic calcium hot zone (Bast and Oberlaender, 2021), which corresponds to the site of dendritic spike generation. Consistent with the idea that evolution selects energy-efficient neuronal phenotypes (Niven and Laughlin, 2008), low expression of potassium channels in distal apical dendrite has been observed before in real neurons (Schaefer et al., 2007). Although the authors did not perform Pareto analysis for the economy-computation trade-off, it is tempting to speculate that the models with a best compromise for the two tasks (dendritic computation and energy-efficiency) would lie close to the Pareto front in parameter space for ion channels (Fig. 7). Future experimental and computational analyses might reveal whether the Pareto front for optimal ion channels in L5 PCs resembles a line segment connecting the two archetypes of dendritic computation and energy-efficiency (or a triangle in case of three tasks, etc.). Indeed, the authors concluded that L5 PCs do not exploit all possible parameter combinations but "select those optimized for energy-efficient active dendritic computations". Interestingly, morphological variability did not seem to play a major role, suggesting that dendritic structure is constrained mainly by optimal wiring (as we mentioned before) and does not greatly affect ion channel parameters.

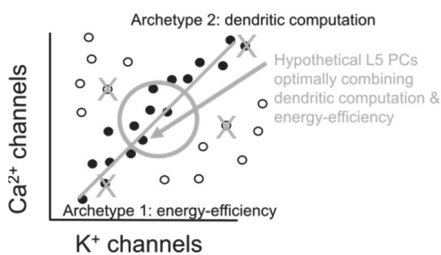

*Figure 7 L5 pyramidal tract neurons (L5 PCs) might be Pareto optimal for a 2-task trade-off between dendritic computation and energy-efficiency. Inspired by population modeling of L5 PCs by Bast and Oberlaender (2021), we hypothesize that Pareto optimal parameter space (i.e. Pareto front) for $K^+$ and $Ca^{2+}$ channels (referring mainly to fast noninactivating potassium channel Kv3.1 and high-/low-voltage activated calcium channels) might be a line. L5 PC models with low Kv3.1 and $Ca^{2+}$ expression (in the dendritic hot zone) compromise optimally between dendritic computation (i.e. dendritic spikes and BAC firing) and low energy costs (Bast and Oberländer, 2021). Grey line: a hypothetical Pareto front with models jointly optimal for combined 2 tasks (filled circles). Grey arrow: a hypothetical model with a relatively low (experimentally observed) expression of Kv3.1 is close to the Pareto front and displays a good combined performance in energy-efficiency (archetype 1) and dendritic computation (archetype 2). Open circles: Pareto non-optimal models. Grey X: hypothetical models removed from parameter space (either non-optimal, or with unrealistically high (low) channel expression and low energy-efficiency (low performance in dendritic computation).*

Another recent study (Deistler et al., 2022) explored how energy-efficiency and temperature robustness affect parameter space of population models for the canonical circuit of the crab stomatogastric ganglion, in which parameter degeneracy was discovered for the first time (Prinz et al., 2004). The authors combined population modeling with a new machine learning method for estimating parameters of mechanistic models (Gonçalves et al., 2020). As expected, energy-efficiency reduced the parameter space for realistically behaving models (Deistler et al., 2022). However, the remaining parameter space was still large and degenerate so that disparate parameter combinations still led to well performing models in terms of energy-efficiency and network behavior. In addition, somewhat surprisingly, increased robustness to temperature did not always cause increased energy consumption. This suggests that in this circuit there might not be a significant trade-off between energy cost and robustness to alterations in temperature. Nevertheless, it does not exclude the possibility that future research employing Pareto optimality theory will discover other tasks (e.g. robustness to other, temperature-unrelated perturbations) that would further constrain parameter space and find stronger trade-offs with energy efficiency. Notably, although the authors did not study Pareto fronts in conductance space, their simulations predict that sodium and calcium conductances contribute significantly to energy costs and are therefore "less variable in nature than expected by computational models only matching network activity". This is in agreement with the hypothesis that evolution does not implement all possible parameter combinations but selects their subsets (see Fig. 2 and 4). Interestingly, the work showed also that individual neuron models could be tuned for low energy costs independently from network activity and then used to construct energy-efficient circuit models. This strengthens the idea that Pareto analysis of economy-function trade-offs can be applied not only to circuits but also to single-cell models of neurons.

**Pareto inference for deducing neuronal functions from high-dimensional Patch-seq data**

Until now we have described possible applications of Pareto theory to known tasks (or functions and their energy costs) of neurons. However, the tasks of most neurons and neural circuits are still not fully understood. Moreover, even if we knew the functions, we might not be able to estimate the associated performance in performance space. Surprisingly, even if no (or not all) functions and corresponding performances of neurons are known, the framework of Pareto optimality can still be used and provide interesting insights.

As mentioned above, in parameter space (but not in performance space), evolutionary multi-task optimization leads to Pareto fronts with specific geometrical shapes (see Fig. 5 for a 2D parameter space). A trade-off between two, three, four or n tasks leads to a Pareto front shaped as a line segment, a triangle, a tetrahedron or a polytope with n vertices and an (n-1) dimensional surface (Shoval et al., 2012; Sheftel et al., 2013; Hart et al., 2015; Alon, 2020). Optimal neuronal phenotypes would be expected to lie inside such polytopes whose vertices represent the archetypes for each task. Remarkably, the theory (Alon, 2020) predicts that Pareto fronts with polytope shapes and sharp vertices will emerge in parameter space independently from the number of measured parameters (i.e. the number of dimensions, see Fig. 8 for a 3D parameter space). For example in two-, three- or higher dimensional parameter space, two tasks always lead to a one-dimensional line segment (a curve) as the corresponding Pareto front. The reason for this is that the projection of the line segment to a plane is again a line segment (Alon, 2020; Fig. 8A). Therefore, in theory, these geometrical shapes (especially their vertices) should be identifiable in experimental datasets irrespective of which or how many parameters were measured (Alon, 2020). Conveniently, if biological data can be fit to the polytopes (i.e. lines, triangles, tetrahedrons etc.) then the sharp vertices or corners can be exploited to infer biological tasks from experimental data (Fig. 8). This innovative approach has been termed Pareto task inference (ParTI, (Shoval et al., 2012; Hart et al., 2015)).

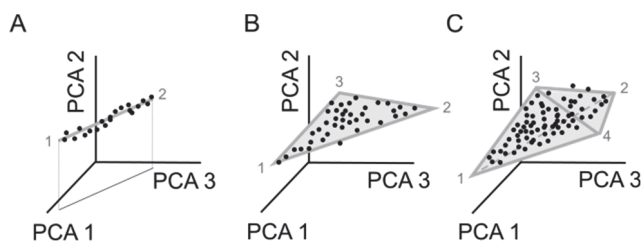

*Figure 8 Pareto task inference (ParTI) may help infer main functions of neurons from single-cell ion channel expression data. High-dimensional datasets, e,g, ion channel expression data from Patch-seq experiments, can be reduced to a 3-dimensional parameter space by principle component analysis (PCA). According to Pareto theory, trade-offs between n tasks will lead to data points filling out geometrical objects (Pareto fronts) with a shape of polytopes with n vertices (2 in A, 3 in B, 4 in C). The vertices (sharp corners) in measured datasets could be used to infer the key tasks of given neuron types as shown previously for nonneuronal cell types (e.g. liver cells – see Adler et al. 2019). However, future research is needed to clarify whether doing PCA first would give the most significant functional archetypes or would lead to a loss in archetypes by projecting to a lower-dimensional space. The figure is based on Alon (2020).*

ParTI has been successfully applied to diverse datasets including morphologies and life history traits in animals (Szekely et al., 2015; Tendler et al., 2015) or parameters in biological homeostatic systems (Szekely et al., 2013) as well as gene expression in bacteria (Shoval et al., 2012), liver cells (Adler et al. 2019) and cancer cells (Hausser et al., 2019; Hausser and Alon, 2020). Motivated by the success of the ParTI even for single-cell data, we propose that ParTI applied to Patch-seq data (see table 1 in Lipovsek et al., 2021) might provide new insights into neuronal functions. Patch-seq experiments generate large amounts of multimodal and high-dimensional data for ion channel expression, electrophysiological behavior and dendrite morphologies of neurons (e.g., Cadwell et al., 2016; Fuzik et al., 2016; Berg et al., 2021; Kalmbach et al., 2021). Unfortunately, higher than three-dimensional datasets are difficult to visualize. However, Pareto optimality makes it plausible that neuronal tasks (such as for example nonlinear dendritic computations, sparse or fast firing, etc. and most likely always also energy-efficiency) simplify the geometrical shape of experimentally observed parameter space of Patch-seq datasets to low-dimensional manifolds. Typically, Patch-seq datasets are analysed by common data clustering analyses. However, clustering analysis presupposes that neuronal data is structured in separate clusters (Alon, 2020) although ion channel degeneracy may lead to data continuity (Scala et al., 2021; see also Bast and Oberlaender, 2021; Schneider et al., 2021). Accordingly, ParTI may be better suitable for a high-dimensional continuum of Patch-seq data (e.g., Scala et al., 2021) than clustering analysis (Alon, 2020). High-dimensionality of datasets is not a great problem since the Pareto optimal geometrical shapes should emerge irrespective of dimensionality. Besides, dimension reduction methods (e.g., principal component analysis, PCA) can help visualize projections of the presumably Pareto optimal shapes in 2D or 3D parameter space (Fig. 8). Interestingly, the above-mentioned population modeling of L5 PCs (Bast and Oberländer 2021) discovered a highly significant correlation of energy-efficiency with the first principle component (PCA1) in hybrid function/parameter space for ion channels (space consisted of total charges flowing through ion channels). Together with non-neuronal cellular examples (Adler et al., 2019; Alon, 2020), this suggests that ParTI (or its improved version, see Adler et al., 2022) might be able to reveal biologically important tasks of analyzed neurons when applied to a high-dimensional ion channel space reduced to three dimensions by PCA. However, it is also possible that PCA will mask some relevant functional archetypes by projecting data to to a lower-dimensional space (Fig. 8). In any case, it will be interesting to test whether ParTI can infer known (or also unknown) functions of neurons from the geometrical shapes of Patch-seq datasets and their sharp corners in parameter space.

Del Giudice and Crespi (2018) have described basic functional trade-offs between four universal tasks of neural systems, namely (1) functional performance (termed "efficiency" in their article, synonymous with "effectiveness" in our article), (2) energy-efficiency (synonymous with "economy"), (3) robustness, and (4) flexibility (see their article for a concise definition of these four universal properties). Converging evidence indicates that trade-offs between the four tasks profoundly shape cognitive, neuronal and synaptic phenotypes (Del Giudice and Crespi, 2018). Correspondingly, it is an exciting question whether these basic four functional properties (or their derivatives) and their trade-offs can be identified, disentangled and clarified with the help of ParTI-analysis of experimental data. Future ParTI-based neuroscience studies might focus on inferring brain region-specific, cell type-specific or/and universal tasks of neurons across brain regions (cf. Hausser et al., 2019).

**Open questions**

Multi-task Pareto optimality is a promising but still largely unexplored framework for studying ion conductance space of neurons and their models. For example, it remains an open question whether Pareto analysis will show that real neurons with their naturally occurring ion channel parameters lie on modeling-based Pareto fronts or not. If data showed that neurons were distant from a predicted Pareto front this could mean that the Pareto analysis did not include the important (i.e. evolutionarily relevant) tasks or that neurons were not close to being Pareto optimal (Pallasdies et al., 2021). In any case, the Pareto framework will provide new testable predictions and insights.

Many other open questions remain to be addressed. For example, if neurons are optimized for the best compromise between function and energy-efficiency, what happens if they face perturbations such as scarcity of energy resources? Interestingly, a recent study (Padamsey et al., 2022) has shown that in animals with food restriction, layer 2/3 pyramidal cells (L2/3 PCs) in visual cortex increase their energy-efficiency (by weakening their input synapses) but reduce their coding precision (as reflected in a broader orientation tuning). However, the firing rate of L2/3 PCs remained unchanged. It would be intriguing to apply Pareto theory to these data. It is tempting to speculate that under food scarcity, neocortical L2/3 PCs moved along the Pareto front closer towards the archetype for maximum energy efficiency but further away from the archetype for the best computational function in the form of visual information processing. Curiously, they still performed well in firing rate homeostasis. Thus, the neurons likely found a new optimal balance between economy and visual computation. It would be interesting to use population modeling and ion channel analyses to find out the shape of the Pareto front in parameter space of L2/3 neurons. It might be a line for a trade-off between economy-visual processing. Alternatively, it might be a triangle if relevant trade-offs include firing rate homeostasis. Or it could be a tetrahedron or another polytope if these cells are optimized for multi-objective trade-offs between more than 3 tasks.

Another exciting and not fully resolved question is whether multi-objective Pareto optimality may provide insights on well-established correlations of ion channels. Ion channel correlations have been observed in experiments (Khorkova and Golowasch, 2007; Schulz et al., 2007; Temporal et al., 2014; Tapia et al., 2018; Iacobas et al., 2019; Kodama et al., 2020) and explored in computational models (Ball et al., 2010; Soofi et al., 2012; Zhao and Golowasch, 2012; O'Leary et al., 2013, 2014). Our hypothesis is that multi-task Pareto optimization of ion channel parameters could shape their homeostatic tuning and lead to ion channel correlations. This is in line for example with the above-mentioned computational prediction that low calcium and potassium channel co-expression appears to be optimal for the trade-off between dendritic computing and low energy cost (Bast and Oberlaender, 2021). Likewise, ion channel expression data from fast-spiking neurons in the vestibular nucleus suggest that co-regulation of ion channels supports optimal balance between high firing rates and their energy-efficiency (Kodama et al., 2020). Other recent computational work using biophysically simple single-compartment models has explored as to when homeostatic co-regulation of ion channels leads to ion channel correlations (Yang et al., 2022). Intriguingly, Yang et al. have addressed this question in the context of ion channel degeneracy and multi-objective optimization. The authors predict that homeostatic ion channel co-regulation can lead to many (degenerate) multi-objective solutions if the number of available ion channels is higher than the number of objectives. For example, more than two ion channels would be required for finding multiple neuronal models with a successful co-regulation of firing rate and energy-efficiency. Thus, homeostatic co-tuning of multiple tasks seems possible only with sufficiently large ion channel diversity (see also Schneider et al., 2021). In addition, ion channel correlations seem to arise if the solution space (defined as a difference between the number of dimensions of parameter space and the number of dimensions of performance space) is low-dimensional (Yang et al., 2022). It would be interesting to compare these predictions to experiments accompanied by simulations in biophysically and morphologically more complex models complemented by Pareto analysis. Pareto theory suggests that irrespective of the dimensionality of parameter space, the Pareto front for n tasks (objectives) is an (n-1) dimensional surface in parameter space (Sheftel et al., 2013).

**Conclusions**

We have seen that Pareto multi-objective optimality is a useful concept for an elegant simplification of the geometry of parameter space. It has been widely employed in engineering, computer science and economics (Censor, 1977; Steuer, 1986; Clímaco et al., 1997; Pardalos et al., 2008). However, only relatively recently has it started being used in molecular biology and in other life science areas (Alon, 2020) including neuroscience (Pallasdies et al., 2021). Importantly, it has been successfully applied not only to phenotypes of organisms (Kavanagh et al., 2013; Tendler et al., 2015) but also to phenotypes of molecules, molecular pathways (Song and Varner, 2009; Savir et al., 2010; Song et al., 2010; Shoval et al., 2012) and cells, including intestinal and liver cells (Adler et al., 2019), cancer cells (Hausser et al., 2019) and nerve cells (Druckmann et al., 2007; Stiefel and Torben-Nielsen, 2014; Remme et al., 2018; Chandrasekhar and Navlakha, 2019).

Therefore, based on these and other examples mentioned here, we believe that Pareto optimality can be fruitfully applied to conductance-based population models of neurons and their circuits, especially if it informs the search for models with optimal trade-offs between economy and neuronal computations and takes advantage of simplifying geometrical rules for Pareto fronts in parameter space (Shoval et al., 2012; Sheftel et al., 2013). In short, we encourage a more frequent usage of Pareto theory and evolutionary economy-effectiveness trade-offs to select optimal and therefore presumably the most realistic neuronal models. Pareto optimality could provide a general conceptual framework to elucidate the diversity in ion channel properties of neurons. This theoretical framework is linked to multi-task evolution theory (Shoval et al., 2012; Alon, 2020) implying that trade-offs between tasks curb ion channel expression to a continuous Pareto front having a shape of a polytope whose vertices represent ion channel expression profiles specializing for a given task (cf. Hausser et al., 2019).